\documentstyle[aps,preprint,epsfig]{revtex}

\begin{document}
\title{Fragment Multiplicity Distributions, a Signal of True Nuclear
Multifragmentation} 
\author{Tarek Gharib\footnote{Permanent address: Ain Shams
University, Faculty of Science, Department of Physics, Abbassia 11566, Cairo,
EGYPT}, Wolfgang Bauer and Scott Pratt} 
\address{Department of Physics and
Astronomy and \\ National Superconducting Cyclotron Laboratory, 
\\ Michigan
State University, East Lansing, MI 48824~~USA} 
\date{\today}
\maketitle

{\abstract Multiplicity fluctuations of intermediate-mass fragments are studied
with the percolation model. It is shown that super-Poissonian fluctuations
occur near the percolation transition and that this behavior is associated with
the fragmentative nature of the percolation model. The consequences of various
choices in defining and binning fragments are also evaluated. Several
suggestions for experiments in nuclear fragmentation are presented. }


The study of nuclear fragmentation at high excitation has proven
enigmatic. Sophisticated models based on statistical
equilibrium\cite{copenhagen,randrup} or mean field simulations that model the
growth of unstable dynamical modes\cite{randrup,chomaz} have been able to
reproduce several features of fragment yields\cite{copenhagen} as measured in
intermediate heavy ion collisions. Even more ambitious microscopic models that
account for the Fermi degeneracy of nuclear matter are currently under
development\cite{feldmeier,ohnishi,kiderlen}. However, simple percolation
models\cite{bauer,campi,stauffer} have perhaps been the most successful in
reproducing fragment yields and their moments over a wide range of
excitations\cite{bauerbotvina95}, and have also had some success in modeling
the fragmentation of atomic clusters\cite{bauerbuckyballs}.

Recently, Moretto and collaborators\cite{moretto} have put forth the
measurement of fragment multiplicity distributions as an insightful tool for
understanding the mechanisms and the driving principles of nuclear
fragmentation. Experimental fragment yields have shown themselves to be well
described by binomial distributions, while the interpretation of the binomial
parameters has been deeply debated\cite{toke,danielewicz,morettoreply}.

Here, we present calculations of fragment multiplicity distributions for
percolation calculations. Our aim is to address the following questions:
\begin{enumerate}
\item Are fragment multiplicity distributions from percolation calculations of
a binomial nature?
\item Is the variance of the multiplicity distribution governed by simple
conservation laws or by other principles.?
\item Do fluctuations near the percolative transition affect fragment 
distributions?
\item In analyzing nuclear experiments, should one bin distributions by
multiplicity, by excitation energy, or by some other criteria?
\end{enumerate}

\begin{figure}

\epsfig{file=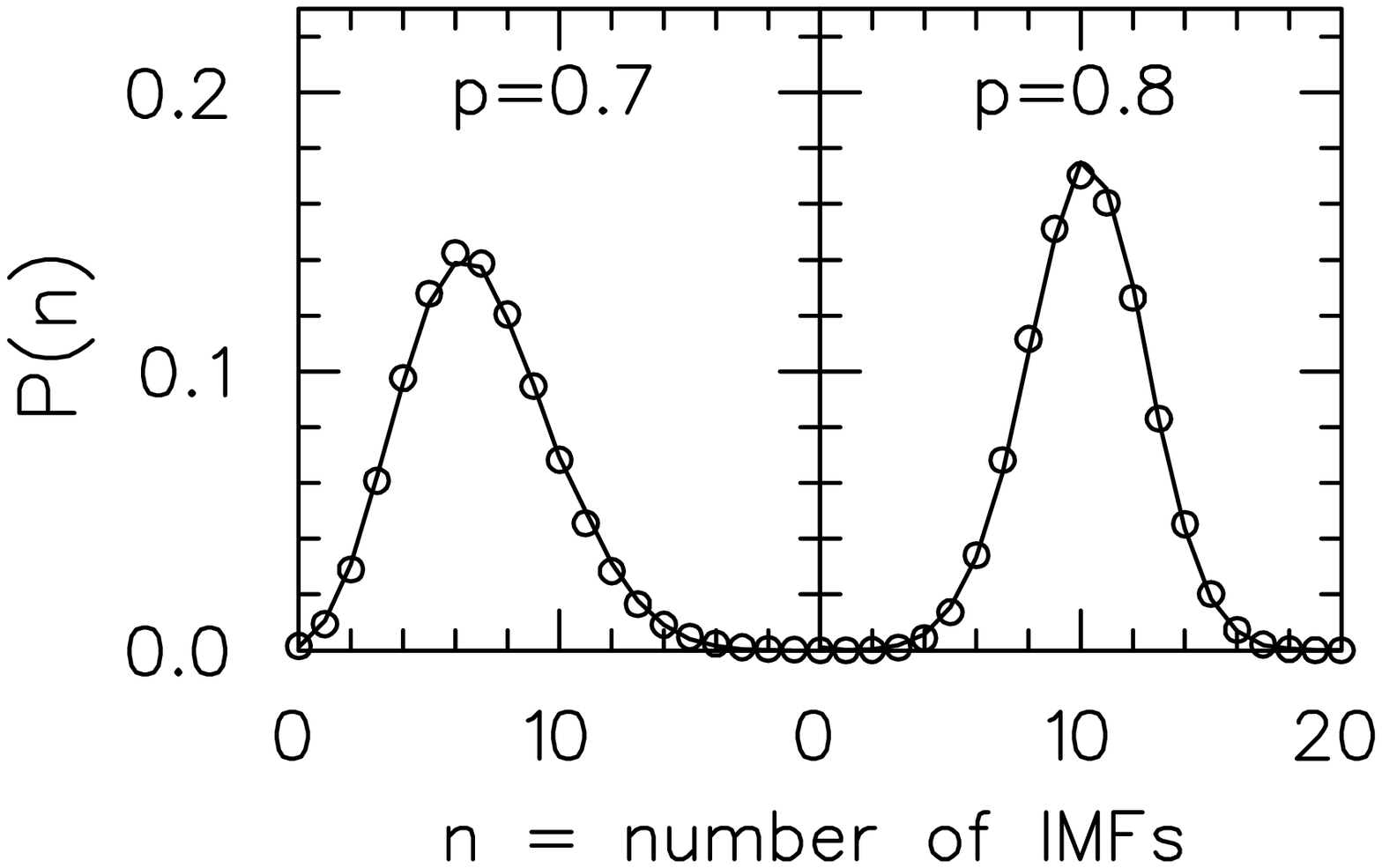,width=7.5cm}\\

\caption{The multiplicity distribution for IMFs for the case of $p$=0.7 (left
panel) and $p$=0.8 (right panel). The lines represent negative binomial (left
panel) and binomial (right panel) fits, where the two parameters were chosen to
match the mean and variance of the distributions.\label{multdis_fig}}
\end{figure}

At first glance, percolation models seem to have little in common with a
nuclear multifragmentative event. No dynamics are present, and bulk properties
such as pressure and specific heat do not even have analogs in a percolative
description. However, percolation models do allow one to study the effects of
particle number conservation and geometry, and therefore can prove insightful
in modeling nuclear fragmentation. In fact, a rigorous connection between bond
breaking probability, deposited energy and nuclear binding energy has been
established\cite{tongli}, which is a generalization of the Coniglio-Klein
formula of the lattice gas model\cite{campikrivine,coniglioklein}. For our
studies we employ bond percolation\cite{bauer,campi,stauffer} where a spherical
section of a cubic lattice is arranged, and bonds between the sites are
randomly broken with a probability $p$. One defines a fragment as a group of
connected sites. For values of $p$ below $p_c=.7512$ the majority of sites
belong to a single large cluster. When $p$ exceeds $p_c$, the lattice is broken
into many small and intermediate sized clusters. We have chosen spherical
lattices of size $N_{sites}$=123 to address lattices of sizes relevant for
nuclear fragmentation and $N_{sites}$=4169 to understand the behavior in larger
lattices. For each event the number $n$ of intermediate-mass fragments (IMFs)
is recorded. The default definition of an IMF is that it is of size,
\begin{equation}
3\le Z \le 20 ,
\end{equation}
where $Z$ refers to the number of sites in the cluster. By recording thousands
of events multiplicity distributions were generated for given values of
$p$. Figure \ref{multdis_fig} displays multiplicity distributions for $p$= 0.7
and $p$=0.8 for the 123-site case.

Moretto and collaborators have reported that the multiplicity
distributions of IMFs in nuclear fragmentation are observed to be well
described by binomial distributions. Binomial distributions are defined by two
parameters $p_b$ and $N_b$,
\begin{equation}
P_b(n)=\frac{N_b!}{n!(N_b-n)!}p_b^n(1-p_b)^{N_b-n}
\end{equation}
The mean and variance of binomial distributions are given by:
\begin{eqnarray}
\label{meanvariance_eq}
\langle n\rangle &=& p_bN_b\\ 
\nonumber
\sigma^2 &=& \langle n\rangle (1-p_b),
\end{eqnarray}
with the variance always being less than the mean.  Thus, by measuring the mean
and variance, one can determine the binomial parameters, $p_b$ and $N_b$. In
the limit that the variance equals the mean the distribution becomes
Poissonian, and if the variance is larger than the mean (super-Poissonian), the
distribution can no longer be considered binomial.  However, one might then
consider the distribution to be a negative binomial,
\begin{equation}
P_{nb}(n)=\frac{(N_{nb}+n-1)!}{(N_{nb}-1)!n!}p_{nb}^n(1-p_{nb})^{N_{nb}}
\end{equation}
In that case the mean and variance become,
\begin{eqnarray}
\left< n\right>  &=& \frac{p_{nb}N_{nb}}{1-p_{nb}}\\
\nonumber
\sigma^2 &=& \frac{\langle n\rangle}{(1-p_b)}
\end{eqnarray}
The lines in Figure \ref{multdis_fig} represent negative binomial and binomial
fits for the $p$=0.7 and the $p$=0.8 cases respectively, where the
parameters were chosen to match the mean and variance of the two
distributions. In all the calculations discussed here, two-parameter fits
were remarkably successful in describing the multiplicity distributions.

\begin{figure}

\epsfig{file=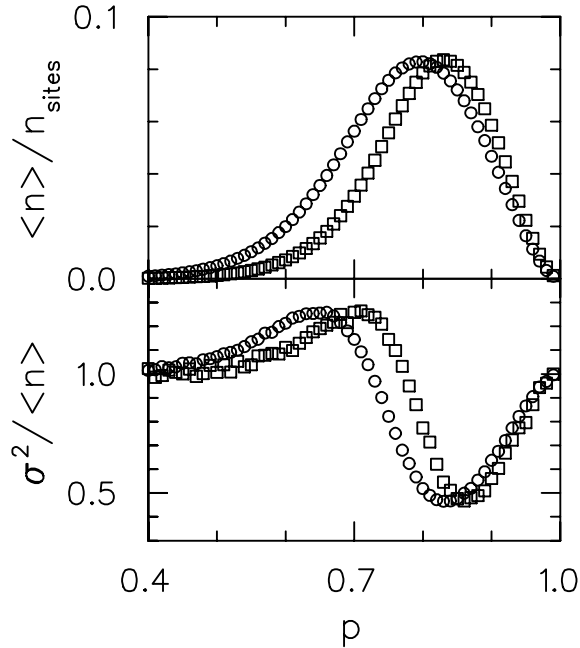,width=7.5cm}\\

\caption{The average multiplicity of IMFs divided by the number of sites, is
displayed in the upper panel. The ratio of the variance to the mean is shown in
the lower panel. A ratio greater than unity is super-Poissonian and signifies
a positive correlation between IMFs. Both 123-site (circles) and the 4169-site
cases are illustrated.\label{nbarsigma2_fig}}
\end{figure}

Figure \ref{nbarsigma2_fig} displays $\langle n\rangle/N_{sites}$ and
$\sigma^2/\langle n\rangle$ as a function of $p$ for the small and large
lattices. The distributions are super-Poissonian for $p<p_c$ and become
sub-Poissonian just above $p_c$. The super-Poissonian behavior is a signal of a
positive correlation between IMFs, as it signals that the presence of an IMF
will be positively correlated with the production of other IMFS. We argue that
this positive correlation is a signal of the fragmentative nature of the
percolation model.

To understand the correlation, we rewrite the expression for the difference of
the variance and mean in terms of a correlation function,
\begin{eqnarray}
\label{correlation_eq}
\sigma^2 -\langle n\rangle
&=& \sum_{a\ne b} \langle(n_a-\bar{n}_a)(n_b-\bar{n}_b)\rangle
+\sum_a \langle n_a^2-n_a-\bar{n}_a^2 \rangle \\
\nonumber
&\approx& \sum_{a\ne b} \langle (n_a-\bar{n}_a)(n_b-\bar{n}_b)\rangle
\end{eqnarray}
The sums over $a$ and $b$ represent the sums over all types of IMFs, where a
type $a$ refers to a specific size, shape and position. The first sum on the
right-hand side of Eq. (\ref{correlation_eq}) represents the correlation
between different IMFs. The second sum can be neglected, as the first two terms
of the second sum cancel each other since $n_a$ can only be zero or unity, and
the last term, which is negative, is small. This last term becomes zero in the
limit that the probability of any specific IMF (defined by size, shape and
position) is small.

Since the bond breaking is random, only fragment types that share the same
sites or the same boundaries are correlated. If type $a$ and type $b$ share any
of the same sites, the correlation is clearly negative as they can not
coexist. This is related to particle number conservation. However, if $a$ and
$b$ merely share some section of their boundaries a positive correlation can
exist. This positive correlation appears only for values of $p$ where a
majority of the sites are taken up by large clusters, larger than the size of
an IMF. The presence of an IMF $a$ then creates extra surface within some
larger cluster. The increased surface area eases the production of a second IMF
of type $b$ which borders the first IMF. As $p$ is increased to the point where
most of the sites are assigned to fragments the same size or smaller than IMFs,
the positive correlation disappears, and the effects of particle-number
conservation are dominant.

The super-Poissonian variance of the IMF multiplicity distribution is a signal
of the fragmentative aspect of the percolation model. The positive correlation
arises from the additional surface created by the production of a fragment. For
sequential models of fragment formation, e.g. an evaporative picture, surface
area is not increased by the production of a fragment, as the nucleus is
assumed to return to it's spherical shape before the production of the next
fragment. In fact, evaporative pictures introduce additional negative
correlations due to energy conservation since the production of an IMF uses a
large amount of energy to surpass the Coulomb barrier, which then makes
subsequent IMF production difficult. Thus, the study of IMF multiplicity
distribution in nuclear collisions provides insight into the general principles
of the fragmentation mechanism.

Several other effects can affect the width of the IMF multiplicity
distribution. One can imagine binning percolation events according to $p$ as
done above, by the number of broken bonds, or by the overall multiplicity. One
might similarly consider binning experimental events by an even larger
assortment of criteria: multiplicity, transverse energy, beam energy, the size
of the largest fragment, or by any combination of the above. One might also
consider altering the mass range that defines an IMF. All such seemingly
arbitrary choices affect the width of the multiplicity distribution. For
nuclear experiments, the choice of criteria is often constrained by details of
the experiment such as acceptance or method of excitation, e.g. symmetric
central collisions of heavy ions at intermediate energy vs. high-energy
peripheral collisions. Thus, we concentrate on the general behavior and
manifestations of various binning criteria in percolation calculations. Many of
the lessons learned from this effort should carry over to the analysis of
nuclear experiments.

Before we proceed, we define a quantity $R(p)$, the difference of the variance
and the mean, normalized by the ratio, $N_{sites}/\langle n\rangle^2$.
\begin{equation}
R(p)\equiv \frac{N_{sites}}{\langle n\rangle^2}\left(\sigma^2-\langle n\rangle\right).
\end{equation}
By dividing by $\langle n\rangle^2$, this normalization allows one to view the
correlation even when fragment production is rare, and by multiplying by
$N_{sites}$, $R$ becomes independent of the lattice size for large lattices.
Positive and negative values of $R$ refer to super and sub-Poissonian
distributions respectively.  

\begin{figure}

\epsfig{file=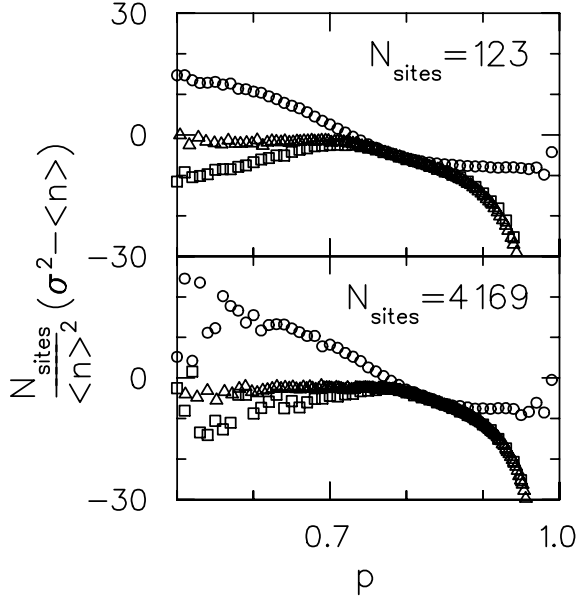,width=7.5cm}\\

\caption{Three different binnings of events are illustrated: Describing the
event by the random probability $p$ of breaking bonds (circles), Categorizing
events by the actual fraction of broken bonds (squares), or binning events
according to the overall multiplicity (triangles). In the last case, the
average value of $p$ for events of a given multiplicity is used to determine
the horizontal axis. For the latter two cases the binnings introduce an
autocorrelation with the number of IMFs that reduces the width of the
multiplicity distribution.\label{differentbinnings_fig}}
\end{figure}

Rather than breaking each bond with probability $p$, one can break a fixed
percentage $p$ of the bonds.  The multiplicity distributions as seen in Figure
\ref{differentbinnings_fig} remain sub-Poissonian for the entire range of
$p$. This extra negative correlation is expected as the presence of an IMF of
type $a$ expended some of the broken bonds. This correlation is non-local as
fragments far away from $a$ are less likely to be produced. This difference
between binning by $p$ and binning by the actual number of broken bonds can be
thought of as being analogous to the difference between the canonical and
microcanonical statistical distributions.

One might also bin events by the overall multiplicity of fragments of any size
rather than $p$. This has the advantage of offering a convenient means of
comparing percolative calculations to experimental results. This also
introduces a negative correlation as the existence of an IMF of type $a$
reduces the net number of other fragments by one, making the existence of a
second IMF less likely. The results of such a binning are displayed in Figure
\ref{differentbinnings_fig}. The binnings were performed by choosing $p$
randomly between 0.4 and 1.0, then binning the event according to multiplicity,
and finally using the average $p$ of events with a given multiplicity as the
horizontal axis. In this way, each point in Figure \ref{differentbinnings_fig}
corresponds to a specific multiplicity but covers a range of $p$ values. Again
the variance is pushed into the sub-Poissonian range. This illustrates the
general principal that any binning criteria that is autocorrelated with the
number of IMFs pushes the IMF multiplicity distribution in sub-Poissonian
direction.

\begin{figure}

\epsfig{file=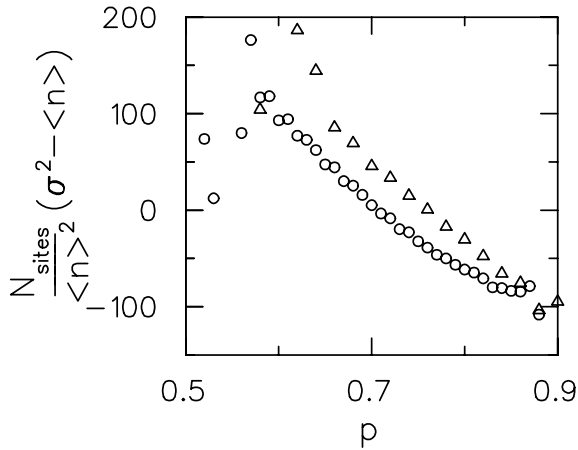,width=7.5cm}\\

\caption{The super-Poissonian nature of the multiplicity distribution is
magnified by restricting the IMF mass range to heavier particles, $15\le Z\le
20$. The 123-site case (circles) and the 4169-site case (triangles) behave
similarly.\label{imfmassrange_fig}}
\end{figure}

We also investigate the effects of changing the mass range that defines an IMF.
The choice of $3\le Z\le 20$ was motivated by a convention for the definition
of IMFs in the analyses of nuclear experiments. By raising the range to $15 \le
Z \le 20$, we see in Figure \ref{imfmassrange_fig} that the multiplicity
distribution becomes increasingly super-Poissonian. In fact, when binning by a
fixed number of broken bonds or a by a fixed multiplicity, the multiplicity
distributions can still be pushed into the super-Poissonian range by using a
larger mass range for IMFs. This positive correlation is due to the fact a
larger fragment offers more surface area for the creation of a second
fragment. Since this positive correlation represents the signal for the
fragmentative nature of the event, we suggest that analyzing more massive
fragments may lead to a more insightful conclusion. However, one must be
careful to steer away from fission-like correlations which set in when the mass
range approaches half the overall lattice.

Finally, we study the consequences of smearing the binning criteria, e.g. the
values of $p$, or some other binning variable, is chosen over a finite range
rather than a discrete value. To understand the effects of a finite range, we
consider a binning variable $x$, where a multiplicity distribution exists for
any discrete value of $x$ with moments $\langle n\rangle(x)$ and $\langle
n^2\rangle(x)$. One can then consider the distribution over a finite range,
$x_{\rm min} \le x\le x_{\rm max}$. Denoting quantities derived from
distributions using the range of $x$ values with overlines, we give expressions
for the difference of the variance and the mean.
\begin{eqnarray}
\overline{\sigma^2}-\overline{\langle n\rangle}&=&
\frac{1}{x_{\rm max}-x_{\rm min}}
\int_{x_{\rm min}}^{x_{\rm max}} dx~ \langle n^2\rangle(x)\\
\nonumber
&&-\left( \frac{1}{x_{\rm max}-x_{\rm min}}
\int_{x_{\rm min}}^{x_{\rm max}} dx~ \langle n\rangle(x) \right)^2
-\frac{1}{x_{\rm max}-x_{\rm min}}
\int_{x_{\rm min}}^{x_{\rm max}} dx~ \langle n\rangle(x).
\end{eqnarray}
Assuming that $\langle n\rangle(x)$ varies linearly in the small range of $x$,
\begin{equation}
\langle n\rangle(x)=\overline{\langle n\rangle}
+\left(x-\frac{x_{\rm max}+x_{\rm min}}{2}\right)
\frac{\langle n\rangle(x_{\rm max})
-\langle n\rangle(x_{\rm min})}{x_{\rm max}-x_{\rm min}},
\end{equation}
one can show that the difference of the variance and mean becomes,
\begin{equation}
\label{smearing_eq}
\overline{\sigma^2}-\overline{\langle n\rangle}=
\frac{1}{x_{\rm max}-x_{\rm min}}
\int_{x_{\rm min}}^{x_{\rm max}} dx~
\sigma^2(x)-\langle n\rangle(x) ~
+\frac{1}{12}\left(\langle n\rangle(x_{\rm max})-\langle n\rangle(x_{\rm min})
\right)^2
\end{equation}
Thus if the distribution is averaged over a region where the average
multiplicity has changed, the resulting distribution is pushed into the
super-Poissonian direction by the last term in Eq. (\ref{smearing_eq}).

Smearing the distribution over a range of $x$ is not necessarily controllable
in a nuclear fragmentation experiment. Due to the inability of an experiment to
gate on a precise type of event such as a central collision, all binnings
effectively cover a finite range of excitation energies. The widening of the
multiplicity distribution is also most affected for regions where the average
IMF multiplicity is rapidly changing as a function of $x$. Clearly, experiments
where all outgoing particles have been measured offer the best chance of
precisely characterizing events and minimizing this effect. 

The behavior observed by Moretto\cite{moretto} is clearly contrary to the
behavior shown in Figure \ref{nbarsigma2_fig}. However, conclusive statements
can not be made without fully understanding the details of the measurement and
analysis. In fact, we are currently investigating systematic effects, such as
the variation of reaction geometry and system size as a function of excitation
energy.

We conclude by answering the four questions posed at the beginning of the
manuscript. First, the multiplicity distributions as predicted from percolation
models can be summarized by two parameters, the binomial parameters for the
cases where the multiplicity distributions are sub-Poissonian and the
negative-binomial parameters for the super-Poissonian cases.

Secondly, the variance of the multiplicity distribution can not be explained
with simple conservation laws. In fact, the most important conclusion from this
study is that measurement of IMF distributions yield an important insight into
the process of fragmentation. If nuclear multifragmentation is truly
fragmentative as in a percolation description, one should expect an increased
variance of the multiplicity distribution, perhaps yielding super-Poissonian
distributions. Furthermore, this variance should increase if one confines the
analysis to larger fragments. Sequential models and thermal models are
expected to behave in the opposite manner, as the presence of a fragment is
anticorrelated to a second fragment due to particle and energy conservation,
and this anticorrelation is increasingly strong for increasingly large
fragments.

The third question centered on the role of fluctuations at the critical
point. The multiplicity distribution at a fixed value of $p$ is not governed by
critical phenomena. The crossover from super-Poissonian to sub-Poissonian
behavior shown in Figure \ref{nbarsigma2_fig} arises from the dissolution of
fragments larger than the IMF size as a function of $p$. Although the fraction
of sites that are part of larger fragments is a rapidly changing function,
there is no associated divergence.

The final question regarding binning resulted in a number of valuable
lessons. Binning by observables such as multiplicity which are correlated to
the number of IMFs narrows the IMF multiplicty distribution, while binning over
a range of excitations effectively broadens the distributions. Thus the fact
that IMF multiplicity distributions are super-Poissonian or sub-Poissonian is
not decisive in itself. But, careful analysis, along with the study of the
behavior as a function of the IMF mass range, should allow one to make
conclusive statements regarding the nature of nuclear multifragmentation. 

During the last 20 years a large number of studies have investigated mass
yields in the region of nuclear multifragmentation, but the shape of the mass
yields has been reproduced by a variety of disparate theoretical
descriptions. The recent development of several full acceptance detector
systems makes the analysis of multiplicity distributions tenable. Using the
percolation model, we have demonstrated that the qualitative nature of
fragmentation might be understood through the analysis of multiplicity
distributions as sequential and percolative descriptions predict qualitatively
different behaviors.

\acknowledgments{This work was supported by the National Science Foundation,
grant 96-05207.}

\end{document}